\documentclass[fleqn]{an}
\usepackage{graphicx,times,fancyhdr}
\usepackage{psfig}
\begin{document}


\def\top{\item}
\def\toptop{\itemitem}
\def\start{\begin{itemize}}
\def\stop{\end{itemize}}
\def\beg{\begin{equation}}
\def\ende{\end{equation}}
\def\Om{{\it \Omega}}
\def\Omst{\Omega^*}
\def\Lam{{\it \Lambda}}
\def\apj{ApJ}
\def\aap{A\&A}
\def\solphys{SoPh}

\title{Differential rotation and  meridional flow in the  solar
       convection zone and beneath}
\author{L.L.~Kitchatinov\inst{1,2} \and G.~R\"udiger\inst{1}}
\institute{Astrophysikalisches Institut Potsdam, An der Sternwarte 16,
           14482, Potsdam, Germany
    \and
           Institute for Solar-Terrestrial Physics, PO Box
           4026, Irkutsk 664033, Russia
}
\date{Received 30 April 2005; accepted 30 May 2005;
published online 1 July 2005}

\abstract{
The influence of the basic rotation on anisotropic and inhomogeneous turbulence is discussed in the context of differential rotation theory. An improved representation for the original turbulence leads to a $\Lam$-effect which complies with the results of 3D numerical simulations.  The resulting rotation law and meridional flow agree well with both the surface observations ($\partial\Om/\partial r<0$ and meridional flow towards the poles) and with the findings of helioseismology. The computed equatorward flow at the bottom of convection zone has an amplitude of about 10 m/s and may be significant for the solar dynamo. The depth of the meridional flow penetration into the radiative zone is proportional to  $\nu^{0.5}_\mathrm{core}$, where $\nu_\mathrm{core}$ is the viscosity beneath the convection zone. The penetration is very small if the tachocline is laminar.
}
\correspondence{lkitchatinov@aip.de}
\maketitle
\section{Motivation}\label{s1}
\begin{figure}[ht]
   \includegraphics[width=9cm,height=8cm]
{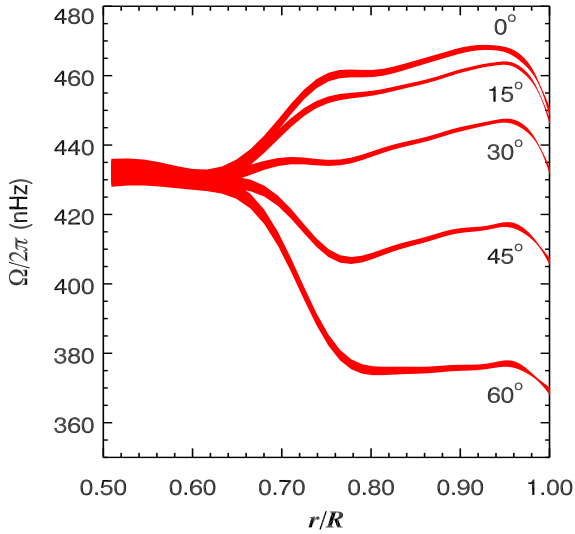}
      \caption{Internal solar rotation from helioseismology.
               Courtesy NSF's National Solar Observatory.
               The numbers give the latitude rather than
               the colatitude.}
      \label{f1}
\end{figure}

The internal solar rotation is well-known from  helioseis\-mo\-logy (Wilson et al. \cite{WBL97}; Kosovichev et al. 1997;  Schou et al. \cite{Sea98}). The decrease of angular velocity with latitude observed on the solar surface survives throughout the convection zone. The differential rotation decreases sharply with depth in the thin \lq tachocline' beneath the convection zone. The  helioseismology inversions also  show a considerable increase of rotation rate with depth just beneath the solar surface  (Fig.~\ref{f1}).

The present paper demonstrates how mean-field models work to explain the overall pattern of the internal solar rotation and to derive the related meridional flow. To this end
\start
\top
our former theory of angular momentum transport by rotating turbulence (Kitchatinov \& R\"udiger \cite{KR93}, hereafter KR93) is extended to the case of slow rotation
\top
the global stellar circulation model is modified to include the meridional flow penetration below the convection zone.
\stop
The currently debated penetration (Nandy \& Choudhury \cite{NC02}; Gilman \& Miesch \cite{GM04}) may be important for advection-dominated models of  the solar dynamo (Choudhuri, Sch\"ussler \& Dikpati \cite{CSD95}; Dikpati \& Gilman \cite{DG01}; Bonanno et al. \cite{Bea02}). The dependence of the penetration depth on basic parameters is computed with our global stellar circulation model. The new simulations include also a computation of the solar tachocline with one of the currently discussed tachocline models (R\"udiger \& Kitchatinov \cite{RK97}; MacGregor \& Charbonneau \cite{MC99}).
\begin{figure}
    \includegraphics[height=4.5cm,width=8cm]{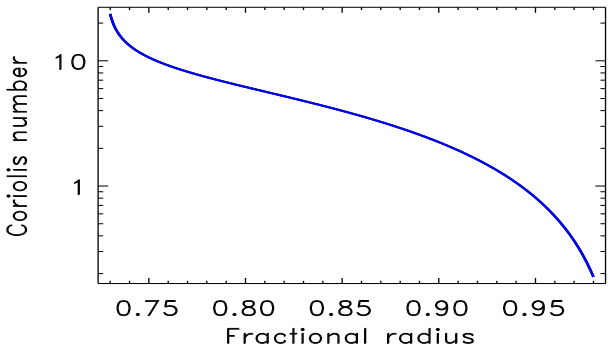}
    \caption{Profile of the Coriolis number (\ref{1}) in the
             solar  model of Stix \& Skaley (\cite{SS90}).}
    \label{Coriolis}
\end{figure}

The angular momentum fluxes inside the convection zone basically depend on  the Coriolis number,
\begin{equation}
 \Omega^* = 2\tau_{\rm corr} \Om ,
 \label{1}
\end{equation}
where $\Om$ is the angular velocity and $\tau_{\rm corr}$ is the convective turnover time. The numerical models (Kitchatinov \& R\"udiger \cite{KR95}, \cite{KR99}; K\"uker \& Stix 2001) based on the $\Lambda$-effect theory of KR93 had practically no free parameters but reproduced the main features of the solar rotation law  and predicted correctly the trend of the surface differential rotation with stellar mass and rotation rate (see Hall 1991; Barnes et al. \cite{Bea05}).

The characteristic value of the Coriolis number for the Sun exceeds unity, $\Omega^*_\odot \simeq 6$, and it is even larger for rapidly rotating stars. The Coriolis number varies, however, with depth inside the Sun (cf. Fig.~\ref{Coriolis}) and becomes rather small  near the surface. The  former models  did not concern  the near-surface region (e.g. Kitchatinov \& R\"udiger \cite{KR95}).

The present paper extends the former  results  by using a more general representation for the spectral tensor of turbulent convection. The resulting expressions for the $\Lam$-effect include a free parameter which is unimportant for $\Omega^* \gg 1$. Its value can be restricted by  numerical experiments. The new model shows agreement with both the observed rotation law and the meridional flow near the solar surface. It predicts the meridional flow towards equator with an amplitude of about 10 m/s at the bottom of the convection zone.
\section{Angular momentum transport}
\subsection{Basic relations}
The ability of turbulence to transport angular momentum is reflected in the structure of the correlation tensor,
\begin{equation}
  Q_{ij} = \langle u'_i(\vec x,t) u'_j(\vec x,t)\rangle ,
  \label{2}
\end{equation}
of the fluctuating velocity ${\vec u}'$. The radial and the latitudinal angular momentum fluxes  are proportional to the off-diagonal components, $Q_{r\phi},\ Q_{\theta\phi}$, of the tensor in spherical coordinates. The fluxes are finite even for the case of rigid rotation; they are conventionally parameterized as
\begin{equation}
  Q^\Lam_{r\phi} = \nu_{\rm T} \Om\ V \sin\theta ,
  \hspace{1 cm}
  Q^\Lam_{\theta\phi} = \nu_{\rm T} \Om\ H \cos\theta .
  \label{3}
\end{equation}
The eddy viscosity, $\nu_{\rm T}$, is introduced here for dimensional reasons, $V$ and $H$ are the normalized vertical and horizontal fluxes.

Negative $V$ also for slow rotation is required to explain the negative radial gradient of the angular velocity  seen in the subsurface layer in Fig.~\ref{f1}. The equatorial acceleration, on the other hand,  demands positive $H$ (R\"udiger 1989). For a nonuniform rotation the correlation tensor (\ref{2}) includes the diffusive part, $Q^\nu$,  which is defined by the eddy viscosity tensor, ${\cal N}_{ijkl}$,
\begin{equation}
  Q_{ij} = Q^\Lam_{ij} + Q^\nu_{ij},
  \label{41}
\end{equation}
with 
\begin{equation}
  Q^\nu_{ij} = -{\cal N}_{ijkl}{\partial \bar{u}_k\over\partial x_l},
  \label{42}
\end{equation}
where $\bar{\vec u}$ is the mean velocity.
\subsection{3D simulations}
It is not easy to separate the $\Lam$-effect from its viscous counterpart in the results of 3D numerical simulations.
K\"apyl\"a et al. (\cite{KKT04}) suggest that the diffusive momentum fluxes may be small in the box simulations.  They  are also reporting $H>0$.

Already Pulkkinen et al. (1993) found that the function $V$ is negative and $H$ is positive. The results of Rieutord et al. (1994) confirmed the positive $H$ but the vertical flux $V$ was small.  Chan (2001) found very clear results also with positive $H$ but with negative $V$. The equatorial value of the latter  proved to be rather small.

Simulations of compressible thermal convection under
the influence of rotation were also made with the finite-difference, fractional-step code NIRVANA (version II) in a small rectangular box defined on a Cartesian grid (Ziegler 1998, 1999; R\"udiger et al. \cite{REZ05}). The Coriolis number (\ref{1}) varied from about unity at the top of the  convection box to about four at its base. The results agree with those of Chan (\cite{C01}) and K\"apyl\"a et al. (\cite{KKT04}) in providing negative $V$ and positive $H$. The ho\-ri\-zontal flux is strongly concentrated towards the equator.

The box simulations  can be summarized as follows.
The radial flux $V$ of the angular momentum is always  negative. Its  equatorial value vanishes for
$\Omega^* > 1$ in agreement with   KR93 for fast rotation. 
The horizontal flux  $H$ is positive in all simulations. It vanishes at the poles as it should be from symmetry reasons.
\subsection{Quasilinear theory}\label{theory}
\begin{figure}
   \includegraphics[height=5.5cm,width=8cm]{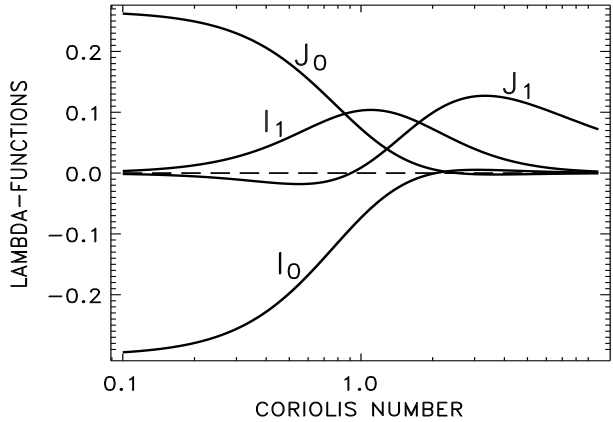}
   \caption{The functions of the Eq.~(\ref{9}).
	    Only $J_0$ and $I_0$ remain finite at small
	    Coriolis numbers and they have opposite signs.
	    For rapid rotation, i.e. large $\Omega^*$, $J_1$
	    dominates all other functions.}
   \label{f4}
\end{figure}

The quasilinear theory  of the angular momentum transport by rotating turbulence in density-stratified fluids results in the  expressions
\begin{eqnarray}
 V &=& V^{(0)}\left(\Omega^*\right)
 - H^{(1)}\left(\Omega^*\right)\ \cos^2\theta ,
 \nonumber \\
 H &=& H^{(1)}\left(\Omega^*\right)\ \sin^2\theta ,
  \label{8}
\end{eqnarray}
for the normalized fluxes\footnote{see the Appendix for more details}. The coefficients $V^{(0)}$ and $H^{(1)}$,
\begin{eqnarray}
  V^{(0)} &=& \left({\ell_{\rm corr}\over H_\rho}\right)^2\left(
  J_0\left(\Omega^*\right) + a I_0\left(\Omega^*\right)\right) ,
  \nonumber \\
  H^{(1)} &=& \left({\ell_{\rm corr}\over H_\rho}\right)^2\left(
  J_1\left(\Omega^*\right) + a I_1\left(\Omega^*\right)\right),
  \label{9}
\end{eqnarray}
depend on the Coriolis number via the functions $J$ and $I$ given by (\ref{a20}) and (A19). The functions $J_0$ and $J_1$ are the same as in KR93. All these \lq$\Lam$-functions' are plotted on Fig.~\ref{f4}. It can be seen from the plots and
Eq.~(\ref{9}) that a positive $H^{(1)}$ requires a slightly positive anisotropy parameter $a$ defined by (\ref{a14}) and (\ref{a17}). On the other hand, a negative  $V$ for slow rotation can be reproduced only with $a > 1$.  The vanishing $V$ at the equator for rapid rotation is due to the  rapid decrease of the functions $J_0$ and $I_0$ with $\Omega^*$.

\begin{figure}
   \resizebox{\hsize}{!}{
   \includegraphics{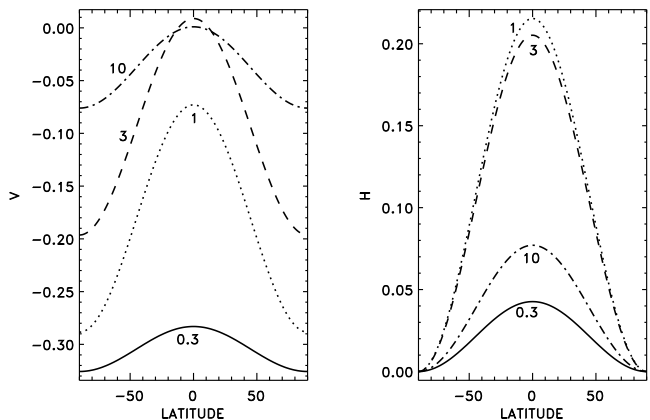}}
   \caption{Normalized fluxes of angular momentum (\ref{8}) in
            dependence on latitude for $a=2$ and $\Omega^*$ from 0.3 to 10.
	    The lines are marked by the correspondent values of the
	    Coriolis number $\Omega^*$. }
   \label{f5}
\end{figure}

Figure~\ref{f5} shows the normalized fluxes, $V$ and $H$,  as functions of latitude for $a=2$, a value that was also used to model differential rotation and meridional flow as discussed below.  The resulting profiles are in agreement with all the main findings of the box  simulations, i.e.
\start
\top
positive $H$,
\top increasingly negative $V$ with increasing latitude,
\top negative equatorial $V$ approaching zero with increasing rotation rate.
\stop

The anisotropy relation
\begin{equation}
\langle u_\phi^{2}\rangle - \langle u_r^{2}\rangle= \frac{1}{8\rho^2}
 \frac{{\rm d^2}}{{\rm d}z^2} \left(\left(1- 2 a/5\right)\rho^2 \ell_{\rm corr}^2 \langle
u^{2}\rangle\right)
\label{ufur}
\end{equation}
follows from (A16) which for $a=0$ yields positive (negative) values in the upper (lower) part of
the solar convection zone. With $a\simeq 2$ the amplitude  of this
anisotropy is strongly reduced in the upper part of the convection zone. In its lower part, the effect is completely quenched by the rapid rotation and the turbulence there remains vertically-dominated.
\section{The model}
The global flow in the solar convection zone was computed with our former model (Kitchatinov \& R\"udiger \cite{KR99}) which now applies the new expressions (\ref{9}) for the $\Lambda$-effect with the functions (\ref{a20}) and (A19). The model solves the steady mean-field momentum  equation,
\begin{equation}
  \left(\bar{\vec{u}}\cdot\nabla\right)\bar{\vec{u}} +
  {1\over\rho}\nabla\cdot\left(\rho Q\right) + {1\over\rho}\nabla P
  = {\vec g} ,
  \label{10}
\end{equation}
for axisymmetric mean flow $\bar{\vec{u}}$ simultaneously  with the entropy equation
\begin{equation}
  \rho T \bar{\vec{u}}\cdot\nabla S +
  {\rm div}\left({\vec F}^{\rm conv} + {\vec F}^{\rm rad}\right) = 0.
  \label{11}
\end{equation}
The entropy equation is needed to define the baroclinic part of the  meridional flow as well as the eddy transport coefficients for the nonrotating case, e.g.,
\begin{equation}
   \nu_{\rm T} = -{\tau_{\rm corr}\ell_{
\rm corr}^2 g\over 15 c_{\rm p}}
   {\partial S\over\partial r} .
   \label{12}
\end{equation}
The model accounts for the anisotropy and quenching of the eddy heat transport coefficients due to global rotation (see R\"udiger et al. 2005a, for details).

The usual stress-free and zero-penetration boundary conditions are  applied. The lower boundary is located at the base of the convection zone (Model~1).  In this model the boundary conditions exclude any penetration of meridional flow in the stably stratified layer beneath the convection zone.
The  magnetic tachocline model by R\"udiger \& Kitchatinov (\cite{RK97}) is also  used in Model~1 so that the rotation law is computed for the entire volume of the Sun. The model realizes the magnetic tachocline as a Hartmann layer due to a weak internal magnetic field of the solar radiative core. The tachocline modeling has no influence on the rotation law computed for the convection zone proper. It only  uses the results of these computations as the top-boundary condition for the tachocline.

Penetration of the meridional flow into the stably stra\-ti\-fied layer beneath the convection zone has been probed in former models but it was always  weak. We  now  rediscuss this issue because the penetration became important in relation with new ideas of solar dynamo theory. The bottom boundary is thus  placed beneath the convection zone inside the stably stratified region (Ekman layer model,  Model~2). For numerical reasons, an extra but small isotropic and uniform viscosity and thermal conductivity were prescribed for the entire computation region. The code is flexible enough to apply a nonuniform grid with about 300 grid points beneath the convection zone to resolve even a shallow  penetration. The dependence of the  penetration depth, $D_{\rm pen}$, on the prescribed effective viscosity ($\nu_{\rm core}$) is computed  with Model~2. The computation of a tachocline makes little sense in combination with this model and it was not attempted.

With the thermal boundary conditions both spherically symmetric heating from below at the bottom of the computational domain and black-body radiation of the photosphere are introduced into the model.
The top boundary of the model is placed shortly below the photosphere to exclude the surface layer with very sharp stratification for which case our representation for the $\Lambda$-effect is certainly incomplete. Nevertheless, in order to include the supergranulation layer into the simulations, the external boundary was shifted from its usual position at  $x_{\rm e} = 0.95$ to  $x_{\rm e} = 0.98$.

Yet, the surface layers are now unstable to large-scale  thermal convection with the usual  value $\ell_{\rm corr} /H_{\rm p} = 1.7$ of the mixing length $\alpha_{\rm MLT}$ (Tuominen et al. \cite{TBMR94}). We had to increase the parameter to $\alpha_{\rm MLT} = 2.1$ to avoid this instabi\-lity. With such a  large value  the mixing length is larger than the density scale height, which is problematic for the construction procedure of the spectral tensor used in the Appendix.

The solar  model of Stix and Skaley (\cite{SS90}) is used. The temperature and density at the external boundary are  $\rho_{\rm e} = 1.37\cdot 10^{-3}$~g~cm$^{-3}$ and  $T_{\rm e} = 1.03\cdot 10^5$\,K,  taken as the  input parameters of the simulations.
\section{Results and discussion}
\label{results}
\subsection{Global flow structure}
After a confrontation of the quasilinear theory of the $\Lam$-effect with the results of the numerical simulations only positive values of the anisotropy $a$-parameter are reasonable. We shall see that for $a>0$ the rotation law in the supergranulation layer well approaches the observed state with $\partial \Om/\partial r <0$ shown in Fig. \ref{f1}. The meridional flow at the surface is poleward and it is equatorward at the bottom of the convection zone. In particular, the resulting amplitude of the flow is an important outcome of the simulations. Doppler measurements by Komm et al. (\cite{KHH93}) show a poleward flow on the surface,  and the helioseismology  indicates that this direction of the circulation persists  to a depth of at least 12~Mm (Zhao \& Kosovichev \cite{ZK04}). A characteristic amplitude for the surface velocity is 10 m/s.

\begin{figure}
   \resizebox{\hsize}{!}{
   \includegraphics{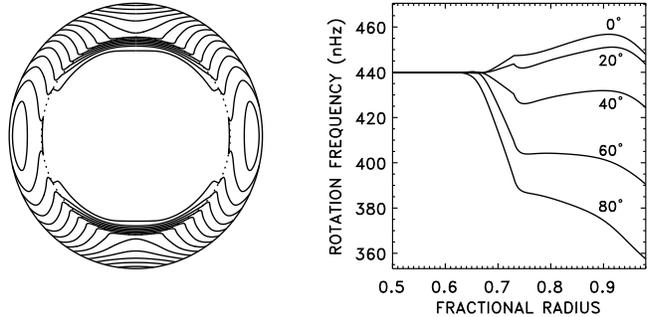}}
   \caption{Rotation law for $a=2$. Left: angular velocity
	    isolines. Right: depth profiles of the rotation rate.}
   \label{f9}
\end{figure}

\begin{figure}
   \resizebox{\hsize}{!}{
   \includegraphics{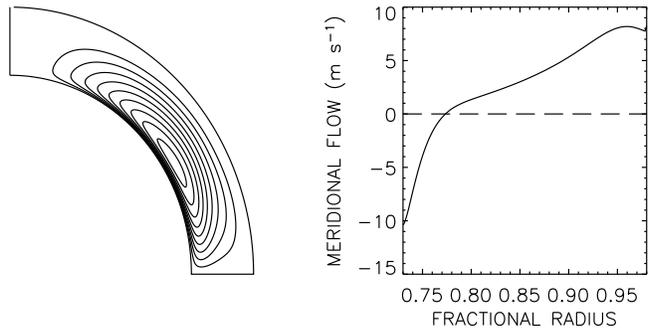}}
   \caption{Streamlines of the meridional flow (left) and flow
            velocity for 45$^\circ$ latitude as function of depth
	    (right) for $a=2$. Positive velocity
	    means poleward flow. One cell of anticlockwise
	    circulation occupies the entire convection zone.}
   \label{f10}
\end{figure}

The global circulation computed with $a=2$ is shown in the Figs.~\ref{f9} and \ref{f10}. The radial gradient of $\Om$ and the direction of meridional flow in the subsurface region are close to the observed state.
The computed rotation law reflects all the known basic features, i.e.
\start
\top  equatorial acceleration,
\top  negative gradients of $\Om$ in the outer region,
\top  small variation with depth for latitudes $\sim 30^\circ$,
\top  shallow tachocline
\stop
(Fig.~\ref{f9}).
The  clear subrotation of the surface region ($\partial \Om/\partial x <0$ for $x > 0.9$) is  mainly related to the radial fluxes of angular momentum. Figure~\ref{f4} shows that the functions $I_0$ and $J_0$ responsible for the radial fluxes are almost completely quenched for $\Omega^* > 2$, which corresponds to $x\leq 0.9$ (Fig.~\ref{Coriolis}). Below this depth the results are thus uninfluenced by the special choice of $a$. The subsurface increase of rotation rate may be important for the solar dynamo (Brandenburg \cite{B05}).

The meridional flow at the surface is poleward. Its amplitude is slightly below 10~m/s. The circulation consists of a single cell with a return equatorward flow at the bottom. The flow reversal occurs rather deep so that the amplitude of the bottom velocity, $\sim$10 m/s, is not small despite of the strong downward density increase.

The poles are warmer than the equator by 2.3 K in the model given in
Figs.~\ref{f9} and \ref{f10} for $a=2$.
Many computations with various $a$-values were made. The models with
$a\simeq 2$ reproduce the observations most closely.

Angular momentum transport by rotating turbulence is known to originate from
anisotropies existing in the turbulence and/or from any inhomogeneity of the turbulent fluid. For fast rotation the $\Lam$-effect produced by an inhomogeneity is much stronger compared to the effect of pure anisotropy. This is probably because the rotation can easily modify the anisotropies,  but not the  stratification. The modification of the spectral tensor introduced in this paper influences the anisotropy with consequences for the $\Lambda$-effect and the resulting rotation law close to the surface. By this  modification the anisotropy  parameter $a$ enters the theory. The para\-meter is fixed using the results of 3D numerical box simulations. Global simulations have made considerable progress recently (Robinson \& Chan \cite{RC01}; Brun \& Toomre \cite{BT02}; Brun, Miesch \& Toomre \cite{BMT04}) but seem to suffer still from the pro\-blem of resolution for the smaller scales.
\subsection{Penetration}
\begin{figure}
   \includegraphics[height=6.5cm,width=7.5cm]{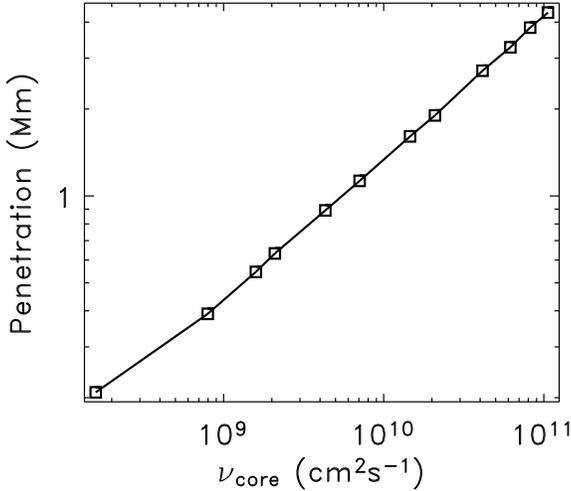}
   \caption{Depth of penetration of the meridional flow into
            the radiative zone at latitude $45^\circ$
            in dependence on the viscosity prescribed
	    for the convectively stable region.
	    The computed dependence is very close to
	    $D_{\rm pen}\propto\sqrt{ \nu_{\rm core}}$.}
   \label{Penet}
\end{figure}

With the Model~2 the penetration depth of the meridional circulation into the radiative zone was computed. The penetrating flow shows multiple reversals of the meridional velo\-city with increasing depth. The velocity amplitude decreases many times after each reversal. The penetration depth $D_{\rm pen}$ of Fig.~\ref{Penet} is defined as the distance from the base of the convection zone to the location of the first reversal of the sign of $\bar{u}_\theta$.

The slope in the plot of Fig.~\ref{Penet} is very close to 0.5. This result complies with the finding of Gilman \& Miesch  (\cite{GM04}) that the penetration under solar conditions belongs to the Ekman regime. The penetration of this type results from viscous drag imposed by the meridional flow at the base of the convection zone on the fluid below. The standard estimate for the Ekman depth is  $D_{\rm pen} \sim \sqrt{ \nu_{\rm core}/2\Omega}$. This is further supported by our finding that the variation of heat conductivity  for constant $\nu_{\rm core}$ does not change $D_{\rm pen}$. The  computations, however, do not reproduce the expected dependence $D_{\rm pen} \propto \Omega^{-0.5}$. The slope is not constant and slightly larger than -0.5. The rotation rate dependence is better (though still not precisely) reproduced by the relation,
\begin{equation}
   D_{\rm pen} \propto \left( \Omega^2 + \Omega {{\rm tan}\theta\over 2}
   {{\rm d}\Omega\over{\rm d}\theta}\right)^{-1/4} ,
\end{equation}
expected for nonuniform rotation.

For large  $\nu_{\rm core}$ the penetration has a considerable effect on the global circulation and the thermodynamics \emph{inside} the convection zone. The meridional flow at the base of the convection zone is reduced and the differential rotation and the differential temperature are increased with allowance for penetration. This is in agreement with Rempel (\cite{R05}) who found that the balance of the meridional forces in the tachocline region demands a  latitudinal entropy gradient which then spreads into the convection zone by radial diffusion. The  poles become even warmer  with this effect and the pole-equator temperature difference resulting from anisotropic diffusion inside the convection zone (R\"udiger et al. \cite{REKK05}) is amplified. Our computations, however, show that the influence of the region of stable stratification on the global flow in  the  convection zone reduces with decreasing $\nu_{\rm core}$ and $\chi_{\rm core}$\footnote{Prandtl number $\nu_{\rm core}/\chi_{\rm core} = 1/8$ in our computations}. For the smallest $\nu_{\rm core}$ given in  Fig.~\ref{Penet} the differential rotation and the temperature distribution inside the convection zone are almost the same as in models without penetration.

Our computations do not support any significant penetration of the meridional circulation beneath the convection zone. The  calculations confirm our former result (Kitchatinov \& R\"udiger \cite{KR95}) that the penetration is very small unless the tachocline below the convection zone is strongly turbulent.
\begin{acknowledgements}
We wish to thank Axel Brandenburg for constructive comments. 
L.L.K. is grateful to the Alexander von Humboldt Foundation and to the Astrophysical Institute Potsdam for hospitality and the visitor support. Russian Foundation for Basic Research is also acknowledged (project 05-02-16326).
\end{acknowledgements}

\appendix
\section{$\Lambda$-effect by rotating inhomogeneous turbulence}
Here we systematically derive the $\Lam$-effect effect for  anelastic turbulent flow of a rotating stratified fluid. The usual strategy to deal with rotating turbulence is to prescribe the turbulence of the nonrotating fluid and then to derive the influence of rotation on the given `original'  turbulence. Here a nonrotating  anelastic flow, ${\rm div}(\rho{\vec u}) = 0$, is considered.

The density stratification is basic for the existence of the $\Lam$-effect.
Fourier modes of the momentum density,
\begin{eqnarray}
 \hat{\vec m}({\vec z},\omega) & = & \int \mathrm{exp} \left(
   -{\rm i}{\vec x}\cdot{\vec z} + {\rm i}\omega t\right)\ \nonumber \\
&     \cdot & {\vec m}\left({\vec x},t\right)\ {\rm d}{\vec x}\ {\rm d}t
   /\left( 2\pi\right)^4,
   \label{a1}
\end{eqnarray}
are used in order to construct the spectral tensor,
\begin{equation}
   \hat{M}_{ij}\left( {\vec z},{\vec z}',\omega , {\omega}'\right) =
   \langle \hat{m}_i\left({\vec z},\omega\right)
   \hat{m}_j\left({\vec z}',{\omega}'\right)
   \rangle ,
   \label{a2}
\end{equation}
of the  fluctuating momentum density, ${\vec m} = \rho{\vec u}$.
The  turbulence is assumed statistically steady. The two new wave vectors,
\begin{equation}
   {\vec k} = \left({\vec z} - {\vec z}'\right) /2,\ \ \ \ \ \ \ \ \ \ \ \ \ \ \
   {\vec\kappa} = {\vec z} + {\vec z}',
   \label{a3}
\end{equation}
are introduced. Then
\begin{equation}
   \hat{M}_{ij}\left( {\vec z},{\vec z}',\omega , {\omega}'\right) =
   \delta\left({\omega}' + \omega\right) \hat{M}_{ij}^{(0)}
   \left({\vec\kappa},{\vec k},\omega\right) .
   \label{a4}
\end{equation}
The dependence on $\vec k$ describes the  properties of the correlations on the small scale while the dependence on $\vec\kappa$ accounts for the turbulence inhomogeneity on large scale (Roberts \& Soward \cite{RS75}). We further assume that the (nonhelical) turbulence does not possess any preferred directions other than its intensity gradient.  The overall structure of the tensor,
\begin{eqnarray}
   \hat{M}_{ij}^{(0)} &=& a\delta_{ij} + b{k_ik_j\over k^2} +
   c {1\over k^2}\left( \kappa_ik_j - \kappa_jk_i\right) +
   d{\kappa_i\kappa_j\over k^2}
   \nonumber\\
   &+&
   e{\left({\vec k}\cdot{\vec\kappa}\right)\over k^4}
   \left(\kappa_ik_j + \kappa_jk_i\right) ,
   \label{a5}
\end{eqnarray}
differs from a former derivation (Kitchatinov \cite{K87}) by the last extra term. The symmetry condition
$
   \hat{M}_{ij}^{(0)}\left({\vec \kappa},{\vec k},\omega\right) =
   \hat{M}_{ji}^{(0)}\left({\vec \kappa},-{\vec k},\omega\right) ,
$
is fulfilled. The divergence-free conditions
\begin{eqnarray}
   \left( k_i + \kappa_i/2\right)\hat{M}_{ij}^{(0)} &=& 0,
   \nonumber \\
   \left( k_j - \kappa_j/2\right)\hat{M}_{ij}^{(0)} &=& 0,
   \label{a7}
\end{eqnarray}
 lead to two vector equations with two nontrivial components each, hence
\begin{eqnarray}
   a + b + c{\kappa^2\over 2k^2} +
   e {\left({\vec k}\cdot{\vec\kappa}\right)^2\over k^4} &=& 0,
   \nonumber \\
   a - 2c + d{\kappa^2\over k^2} +
   e {\left({\vec k}\cdot{\vec\kappa}\right)^2\over k^4} &=& 0,
   \nonumber \\
   b + 2c + e{\kappa^2\over k^2} &=& 0 ,
   \nonumber \\
   c - 2d - 2e &=& 0.
   \label{a8}
\end{eqnarray}
The system is of rank 3 so that three unknowns can be expressed in terms of the other two  (say) $c$ and $e$. It is convenient to express $c$ and $e$ in terms of the spectra $\hat{E}$ and $\hat{E}_1$, i.e.
\begin{eqnarray}
   c &=& {\hat{E}\left({\vec\kappa},k,\omega\right)
   - \hat{E}_1\left({\vec\kappa},k,\omega\right)
   \left({\kappa^2\over 2 k^2} -
   {\left({\vec k}\cdot{\vec\kappa}\right)^2\over 2k^4}\right)
   \over
   32\pi k^2\left( 1 - {\kappa^2\over 4 k^2}\right)},
   \nonumber \\
   e &=& {\hat{E}_1\left({\vec\kappa},k,\omega\right)
   \over
   16\pi k^2}.
   \label{a11}
\end{eqnarray}
This leads to 
\begin{eqnarray}
   \hat{M}_{ij}^{(0)} &=& {\hat{E}
   - \hat{E}_1\left({\kappa^2\over 2 k^2} -
   {\left({\vec k}\cdot{\vec\kappa}\right)^2\over 2 k^4}\right)
   \over
   16\pi k^2\left( 1 - {\kappa^2\over 4 k^2}\right)}\times
   \nonumber\\
& \times  &\left(\left( 1 -{\kappa^2\over 4k^2}\right)\delta_{ij} -
   {k_ik_j\over k^2}\ + \right.  \nonumber\\
& + &
\left.
 {1\over 2k^2}\left(\kappa_ik_j - \kappa_jk_i\right)
   + {\kappa_i\kappa_j\over4 k^2}\right)\ + \nonumber \\
& + & {\hat{E}_1\over 16\pi k^2} \left(\ {\kappa^2\over
   k^2}\left(\delta_{ij}
   - {k_ik_j\over k^2}\right) -
   {\left({\vec k}\cdot{\vec\kappa}\right)^2\over k^4} \delta_{ij}\right.
   \ + \nonumber \\
  & +  & 
   \left. 
   {\left({\vec k}\cdot{\vec\kappa}\right)\over k^4}
   \left(\kappa_ik_j + \kappa_jk_i\right) - {\kappa_i\kappa_j\over k^2}
   \right) .
   \label{a12}
\end{eqnarray}
It is further assumed that $\hat{E}$ and $\hat{E}_1$ depend on the wave number $k$ only rather than on the wave vector $\vec k$. The function $\hat{E}$ is the spectrum of turbulence intensity,
\begin{equation}
   \langle m^2\left({\vec x}\right)\rangle = \rho^2\langle u^2\rangle =
   \int\limits_0^\infty \int\limits_0^\infty E\left({\vec x},
   k,\omega\right)\
   {\rm d}k\ {\rm d}\omega,
 \label{a13}
\end{equation}
with
\begin{equation}
 E\left({\vec x}, k,\omega\right) = \int
   {\rm exp}\left({\rm i}{\vec x}\cdot{\vec\kappa}\right)
   \hat{E}\left({\vec\kappa}, k,\omega\right)\ {\rm d}{\vec\kappa} .
   \label{a13a}
\end{equation}
The spectrum  $E_1$  does not contribute to the turbulence intensity but it controls the anisotropy,
$\langle u^2_r\rangle/\langle u^2_\theta\rangle$. The dimensionless anisotropy parameter, $a$, can be written as
\begin{equation}
  a = \frac{4}{\langle m^2\left({\vec x}\right)\rangle}
  \int\limits_0^\infty\int\limits_0^\infty
  E_1\left({\vec x}, k,\omega\right){\rm d}k {\rm d}\omega ,
  \label{a14}
\end{equation}
and it can depend on position.
For the special case of
\start
\top   $\hat{E}_1/\hat{E} = \mathrm{const} = a/4$,
\top  $E\propto\delta\left( k - \ell_{\rm corr}^{-1}\right)$  (mixing-length approximation),
\top  $\nabla\langle m^2\rangle = -2{\vec g}\langle m^2\rangle/L$ ($\vec g$ is radial unit vector and $L$ is a constant length),
\stop
the spectral tensor (\ref{a12}) can be integrated over wave space to give the one-point correlation tensor, i.e.
\begin{eqnarray}
   Q_{ij} &=& {\langle u^2\rangle\over 1 + \ell_{\rm corr}^2/L^2}
   \left({1\over 3}\delta_{ij} + {\ell_{\rm corr}^2\over 2 L^2}
   \left(\delta_{ij} - g_ig_j\right)\right) ,
   \nonumber \\
   &-& {a\langle u^2\rangle\over 1 + \ell_{\rm corr}^2/L^2}
   {\ell_{\rm corr}^2\over L^2}{1\over 15}\left(\delta_{ij} - 3g_ig_j\right) .
   \label{a15}
\end{eqnarray}
For the rms intensities in the  radial  and the horizontal directions it yields
\begin{eqnarray}
  \langle u^2_r\rangle &=& {\langle u^2\rangle\over 1 + \ell_{\rm corr}^2/L^2}
  {1\over 3}\left( 1 + {\ell_{\rm corr}^2\over L^2}{2 a\over 5}\right) ,
  \\
  \langle u^2_\theta\rangle &=& \langle u^2_\phi\rangle =
  {\langle u^2\rangle\over 1 + \ell_{\rm corr}^2/L^2}
  {1\over 3}\left( 1 + {3\ell_{\rm corr}^2\over 2 L^2}-
  {\ell_{\rm corr}^2\over L^2}{a\over 5}\right) .
 \nonumber 
\label{a16}
\end{eqnarray}
The condition that the turbulence intensities (\ref{a16}) are  positive-definite restricts the value of the anisotropy parameter $a$,
\begin{equation}
  -{5 L^2\over 2 \ell_{\rm corr}^2} \leq a\leq {15\over 2} + {5L^2\over \ell_{\rm corr}^2} .
  \label{a17}
\end{equation}
Isotropy ($\langle u^2_r\rangle = \langle u^2_\theta\rangle$) results for  $a = 5/2$. Although the turbulence is isotropic in this case, the rotation produces a  $\Lam$-effect with negative $Q_{r\phi}$.

The ratio $\kappa/k$ in Eq.~(\ref{a12}) is of  order of
$\ell_{\rm corr} /L$. For a weakly inhomogeneous turbulence, therfore,  the terms of higher than second order in the scale ratio can be neglected so that 
\begin{eqnarray}
   \hat{M}^{(0)}_{ij} & = &
   {\hat{E}\left({\vec\kappa},k,\omega\right)\over
   16\pi k^2}\left( \delta_{ij} - \left( 1 + {\kappa^2\over
   4k^2}\right) {k_ik_j\over k^2} \right.
   \nonumber\\
   & + &  \left. 
   {1\over 2 k^2}\left( \kappa_i k_j - \kappa_jk_i\right)
   + {\kappa_i\kappa_j\over 4 k^2}\right)
   \\
   &+& {\hat{E}_1\left({\vec\kappa},k,\omega\right)
   \over 16 \pi k^4 }
   \left({\left({\vec k}\cdot{\vec\kappa}\right)\over k^2}
   \left(\kappa_ik_j + \kappa_jk_i\right)
   - {\left({\vec k}\cdot{\vec\kappa}\right)^2\over k^2}\delta_{ij}
   \right.
   \nonumber \\
   & - & \left. \kappa_i\kappa_j  + {1\over 2} \left( \kappa^2 +
   {\left({\vec k}\cdot{\vec\kappa}\right)^2\over k^2}\right)
   \left(\delta_{ij} - k_ik_j/ k^2\right) \right)
 \nonumber  
\label{a18}
\end{eqnarray}
results. This equation directly leads to the $\Lambda$-effect.
After the standard derivations (KR93), one arrives at 
\begin{eqnarray}
   Q_{ij}^\Lam &=& \nu_{\rm T}\ \Om_k g_l
   \left( V^{(0)}\left(\Omega^*\right)
   \left( g_i\epsilon_{jkl} + g_j\epsilon_{ikl}\right)
   \right.
    \nonumber \\
   &-& \left. H^{(1)}\left(\Omega^*\right)
   {\left({\vec g}\cdot{\vec\Om}\right)\over\Om^2}
   \left( \Om_i\epsilon_{jkl} + \Om_j\epsilon_{ikl}\right)
   \right) .
   \label{a19}
\end{eqnarray}
The azimuthal  components of this tensor reproduce Eq.~(\ref{3}) with the normalized fluxes of angular momentum defined by (\ref{8}) and (\ref{9}).
The functions $J_0$ and $J_1$ (\ref{9}) are the same  as in KR93,

\newpage
\begin{eqnarray}
   J_0\left(\Omega^*\right) &=& {1\over 2 {\Omega^*}^4} \left(
   9 - {2 {\Omega^*}^2\over 1 + {\Omega^*}^2} -
   {{\Omega^*}^2 + 9\over{\Omega^*}}\ {\rm arctan}\ {\Omega^*}\right) ,
   \nonumber \\
   J_1\left(\Omega^*\right) &=& {1\over 2 {\Omega^*}^4} \left(
   45 + {\Omega^*}^2 - {4{\Omega^*}^2\over 1 + {\Omega^*}^2} \right.
   \nonumber \\
   &+& \left. {{\Omega^*}^4 - 12{\Omega^*}^2 - 45\over{\Omega^*}}\
   {\rm arctan}\ {\Omega^*}\right) ,
   \label{a20}
\end{eqnarray}
and the expressions
\begin{eqnarray}
   I_0\left({\Omega^*}\right) &=& {1\over 4{\Omega^*}^4} 
\nonumber\\
& \times &
\left(
   -19 - {5\over 1 + {\Omega^*}^2}
   + {3 {\Omega^*}^2 + 24\over {\Omega^*}}\
   {\rm arctan}\ {\Omega^*}\right),
   \nonumber \\
   I_1\left({\Omega^*}\right) &=& {3\over 4{\Omega^*}^4} 
\\
& \times &
\left(
   -15 + {2 {\Omega^*}^2\over 1 + {\Omega^*}^2} + 
  {3{\Omega^*}^2 + 15\over{\Omega^*}} {\rm arctan} {\Omega^*}
   \right)
 \nonumber
  \label{a21}
\end{eqnarray}
are new. For  slow rotation (${\Omega^*} \ll 1$) only the radial flux of angular momentum exists, i.e.
\begin{equation}
   J_0 \simeq {4\over 15},\ \ I_0 \simeq -{3\over 10},\ \
   J_1 \simeq I_1 \simeq O\left({\Omega^*}^2\right) .
   \label{a22}
\end{equation}
Note that $I_0$ is negative for slow rotation. For fast rotation (${\Omega^*} \gg 1$), $J_1$ dominates  all other functions,
\begin{equation}
   J_1 \simeq {\pi\over 4{\Omega^*}},\ \ J_0 \simeq I_0 \simeq I_1
   \simeq O\left({\Omega^*}^{-3}\right) .
   \label{a23}
\end{equation}
The contribution of the anisotropy parameter $a$ vanishes for fast rotation.

\end{document}